\pdfoutput=1
\documentclass[10pt,conference]{IEEEtran}
\usepackage{amsmath,amssymb,amsfonts}
\usepackage{graphicx}
\usepackage{textcomp}
\usepackage{xcolor}
\usepackage{booktabs}
\usepackage{multirow}
\usepackage{balance}
\usepackage{listings}
\usepackage{url}
\usepackage{xspace}
\usepackage{enumitem}
\usepackage{colortbl}
\usepackage{cite}

\usepackage[numbers,sort&compress]{natbib}

\definecolor{g1}{RGB}{119,221,119}
\definecolor{g2}{RGB}{223,255,0}
\definecolor{g3}{RGB}{255,253,208}

\setlist{topsep=2pt,itemsep=1pt,parsep=0pt}
\graphicspath{{figs/}{./}}

\newcommand{\method}{\textsc{Caret}\xspace}
\newcommand{\eg}{e.g.,\xspace}
\newcommand{\ie}{i.e.,\xspace}

\makeatletter
\newcommand{\linebreakand}{%
  \end{@IEEEauthorhalign}
  \hfill\mbox{}\par
  \mbox{}\hfill\begin{@IEEEauthorhalign}
}
\makeatother

\begin{document}

\title{\method: Training-Free Test-Time Scaling for\\Repository-Level Code Completion}

\author{\IEEEauthorblockN{Jiajie Wang}
\IEEEauthorblockA{\textit{The University of Arizona}\\
jiajiewang@arizona.edu}
\and
\IEEEauthorblockN{Yutong Zhao}
\IEEEauthorblockA{\textit{California State University, Long Beach}\\
yutong.zhao@csulb.edu}
\and
\IEEEauthorblockN{Kebin Peng}
\IEEEauthorblockA{\textit{East Carolina University}\\
pengk24@ecu.edu}
\linebreakand
\IEEEauthorblockN{Sen He}
\IEEEauthorblockA{\textit{The University of Arizona}\\
senhe@arizona.edu}
\and
\IEEEauthorblockN{Qing Guo}
\IEEEauthorblockA{\textit{Nankai University}\\
qing.guo@nankai.edu.cn}
\and
\IEEEauthorblockN{Tianlin Li}
\IEEEauthorblockA{\textit{Nanyang Technological University, Singapore}\\
tianlin001@e.ntu.edu.sg}}

\maketitle

\begin{abstract}
Retrieval-augmented generation (RAG) dominates repository-level code completion: it retrieves cross-file context (R), then decodes one greedy completion (G).
Existing work mainly focuses on retrieval and stops there.
We argue both stages can be improved together, with generation in particular gaining from test-time scaling.
We present \method{}, a training-free method. For R, \method{} routes among retrieval contexts using the agreement among its own samples, cascading to an alternative context when the samples scatter. For G, it samples candidates over a cached prompt prefix, so the long retrieved context is encoded once rather than once per sample. It then selects the final completion by reverse-context likelihood: a correct completion makes the code after the cursor more probable, so the same model grades its own candidates by reading ahead. Across CrossCodeEval, RepoEval-Line, and RepoEval-API with six code models (1.1B to 7B, four families), \method{} improves exact match in all 18 combinations by 10.8 points on average over greedy decoding and 5.3 over self-consistency@10. Token-level compute stays near 1.45 times one generation (measured wall-clock 1.3--2.9 times, growing with generator size). Improving retrieval and generation together yields more accurate code than improving retrieval alone, at a budget that stays close to a single pass.
\end{abstract}


\section{Introduction}\label{sec:intro}

\noindent Repository-level code completion asks a language model to complete code in a file using knowledge of the whole repository; we study its most frequent granularity, the next line or API call at the cursor. The correct answer often depends on identifiers, signatures, and call patterns defined elsewhere in the project~\cite{ding2023crosscodeeval,zhang2023repocoder,liu2024repobench}. A model that sees only the current file misses these cross-file dependencies. Missing context, rather than limited model capacity, is the dominant failure mode~\cite{ding2023crosscodeeval}. The central problem is therefore supplying the right evidence from the repository before the model decodes.

Current approaches to repository-level code completion mainly concentrate on what to retrieve. Retrieval-augmented generation (RAG) assembles cross-file evidence into the prompt through iterative lexical retrieval~\cite{zhang2023repocoder}, dataflow analysis~\cite{cheng2024draco}, code graphs~\cite{liu2024graphcoder}, learned retrievers~\cite{wang2024rlcoder}, and multi-source fusion~\cite{liang2024repofuse,coderag2025}, then decode once with greedy search. However, \textbf{whole-pipeline optimization from retrieval to generation remains underexplored}, especially the generation stage. This motivation is further supported by our motivating study in Section~\ref{sec:bg-motivation}.

\begin{figure}[t]
  \centering
  \includegraphics[width=0.95\columnwidth]{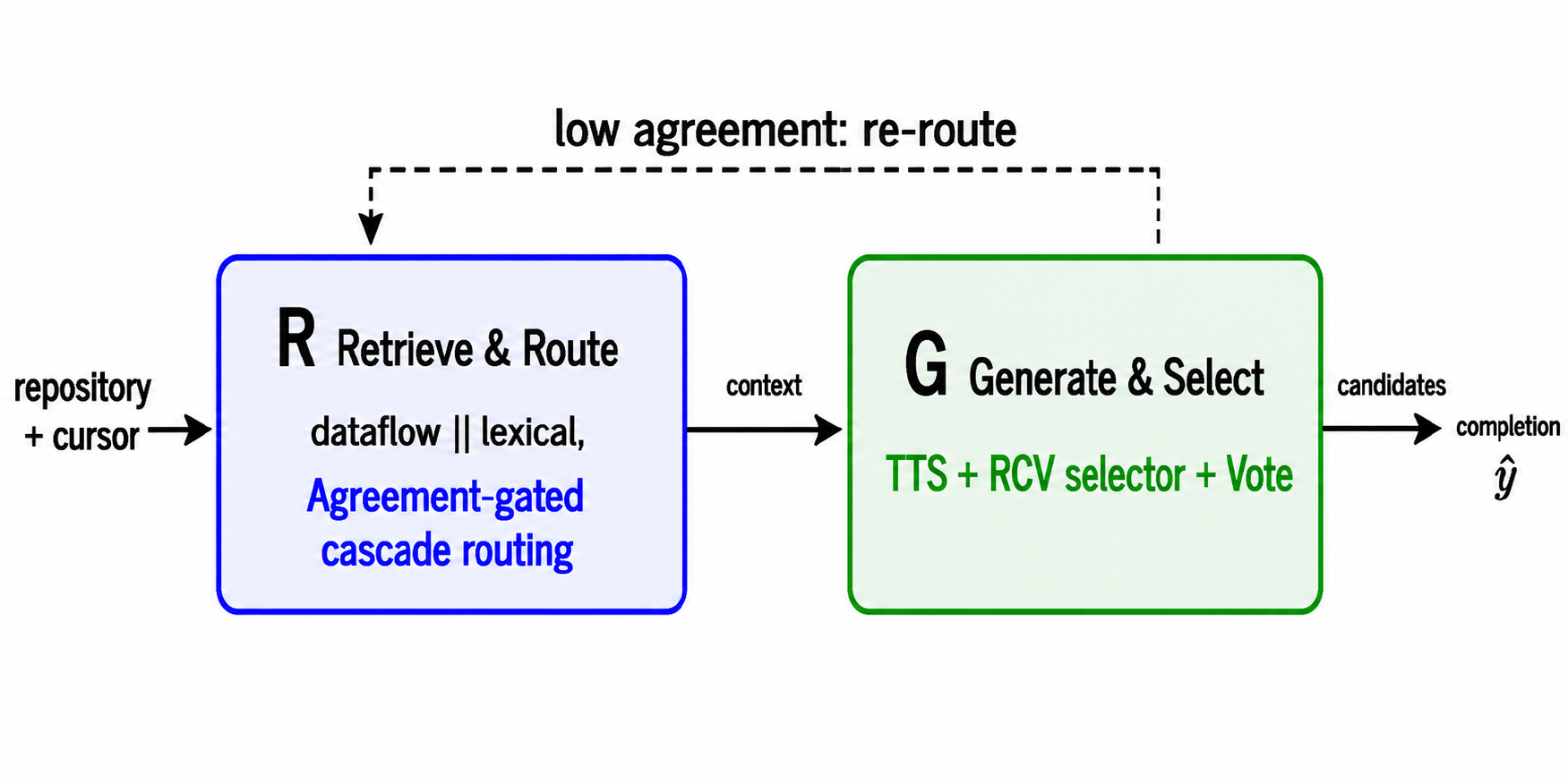}
  \caption{The \method{} pass. Retrieval and routing (R) pick a context using sample agreement; generation (G) samples candidates over a cached prefix; a final selection step reranks distinct candidates by reverse-context likelihood and vote frequency.}
  \label{fig:flow}
\end{figure}

To address this, we propose \method{}, which improves both stages of the pipeline under one inference budget. Figure~\ref{fig:flow} sketches the pass. For \textbf{retrieval and routing (R)}, \method{} reads the agreement among its own samples (the fraction matching the modal candidate) and re-samples from an alternative retrieval context only when agreement is low. For \textbf{generation (G)}, it scales test-time compute by sampling over a cached prompt prefix, which encodes the long retrieved context once rather than once per sample, and reranks the distinct candidates by their reverse-context likelihood blended with vote frequency, skipping scoring when agreement is already high. The same small model performs every step, with no second model, training, or labels.

The central problem is to raise accuracy while keeping the cost increase within a reasonable range. Sampling many candidates and scoring them is the standard way to raise accuracy~\cite{wang2023selfconsistency,brown2024monkeys}, but each extra sample re-encodes the prompt and verifier-based selection adds a scoring pass per candidate~\cite{cobbe2021verifiers,ni2023lever,zhang2023coderreviewer}; on the long repository prompts here, naive sampling multiplies an already heavy cost, and reported multi-pass methods reach up to 1.9 times one generation (Section~\ref{sec:background}). \method{}'s cached prefix and skip-and-deduplicate rules hold this increase to a small factor, and Section~\ref{sec:background} shows that test-time selection already helps without any retrieval, at little extra cost.

Thorough evaluation shows that \method{} improves all 18 model--benchmark combinations over greedy decoding and over self-consistency@10~\cite{wang2023selfconsistency}, by $+10.8$ and $+5.3$ EM on average, the best accuracy among the reproduced baselines under a common prompting and budget protocol. Our strongest generator, DeepSeek-Coder-6.7B, reaches 47.74 EM on RepoEval-API and 43.88 on RepoEval-Line with \method{}, above every retrieval baseline by $+7.6$ and $+9.4$ EM. We claim that a small, training-free test-time budget spent across both stages of the pipeline (retrieval with routing, and generation with sampling and selection) yields more accurate code than improving retrieval alone, and does so for every generator we tried.

We claim the following contributions.
\begin{itemize}
\item We developed \method{}, a training-free method that optimizes the whole retrieval-augmented completion pipeline, both retrieval and generation, without incurring significant extra cost.
\item For retrieval, we develop agreement-gated cascade routing across retrieval contexts (R), which reuses sample agreement to commit to one context or fall through to an alternative, recovering complementarity that a fixed choice forfeits.
\item For generation, we develop cache-aware sampling (G) that encodes the long context once, and the reverse-context verification (RCV) selector that turns the code after the cursor into a verifier, with skip-and-deduplicate rules that bound cost.
\item Thorough evaluation on three benchmarks (CrossCodeEval, RepoEval-Line, RepoEval-API) and six LLMs (1.1B to 7B, four families), where \method{} reaches the best budget-conscious accuracy in all 18 combinations among the reproduced baselines under a common protocol, with a full replication package released upon acceptance.
\end{itemize}

The rest of this paper is organized as follows.
Section~II covers background, related work, and the motivating study, showing why retrieval-only optimization is insufficient and why test-time selection can improve repository-level code completion.
Section~III presents the design of \method{}, including agreement-gated retrieval routing, cache-aware candidate generation, reverse-context verification, and the skip-and-deduplicate cost controls.
Section~IV describes the experimental setup, including benchmarks, models, baselines, metrics, and implementation details.
Section~V reports the evaluation results across CrossCodeEval, RepoEval-Line, and RepoEval-API, and analyzes the accuracy, cost, and ablation results of \method{}.
Section~VI discusses threats to validity and the broader implications of the results.
Section~VII concludes the paper and outlines future work.

\section{Background and Related Work}
\label{sec:background}

\subsection{Repository-Level Code Completion}
\label{sec:bg-task}

\noindent Repository-level completion is defined by where the evidence comes from (other files of the repository), not by how much is completed. We study its line and API-invocation granularity~\cite{zhang2023repocoder,ding2023crosscodeeval}: the completion an IDE offers inline at every pause, where the request rate is highest and the latency budget tightest~\cite{peng2023copilot}, and where a single reference line is a reasonable proxy for correctness because a statement admits far fewer equivalent forms than a function body. Function-level completion is a different regime, with longer outputs, weaker agreement, and execution-based metrics; we return to it in \S\ref{sec:discussion}. A task instance is a tuple $(X_l, X_r, \mathcal{R}, y)$. The \emph{left context} $X_l$ precedes the completion point in the current file, the \emph{right context} $X_r$ follows it, $\mathcal{R}$ is the set of remaining repository files, and $y$ is the reference completion. A generator $\mathcal{M}$, a pretrained code language model~\cite{chen2021codex,guo2024deepseekcoder,nijkamp2023codegen}, produces a candidate $\hat{y}$. Correct completions often hinge on identifiers, signatures, and usage patterns defined in \emph{other} files, so prompting with the current file alone performs poorly~\cite{ding2023crosscodeeval,liu2024repobench}. The standard remedy is retrieval augmentation: a retriever maps $(X_l,\mathcal{R})$ to a token-budgeted \emph{base context} $C$, and the completion is decoded as $\hat{y}\sim\mathcal{M}(\cdot \mid C \oplus X_l)$, where $\oplus$ denotes prompt concatenation.

Three public benchmarks instantiate this task, each instance shipping all four tuple components. CrossCodeEval (Python) contains 2,124 line-completion tasks constructed so the target needs cross-file information~\cite{ding2023crosscodeeval}. RepoEval-Line contains 1,600 line-completion tasks, and RepoEval-API 687 API-invocation tasks over 8 repositories~\cite{zhang2023repocoder,wu2024repoformer}. The primary metric is exact match (EM) of the first non-comment generated line against the reference; edit similarity (ES)~\cite{lu2021codexglue} is secondary.

One property of the setting recurs below. The right context $X_r$ is available at completion time: mid-file completion is the common case in these benchmarks and in fill-in-the-middle training~\cite{bavarian2022fim}, and every instance ships $X_r$. Under the causal prompting we adopt for all six generators, $X_r$ never enters the prompt: an available but unused signal.

\subsection{Retrieval-Augmented Generation}
\label{sec:bg-rag}

\noindent Repository-level retrieval-augmented generation (RAG) improves what goes into the base context $C$. RepoCoder retrieves iteratively, feeding a draft completion back into the next retrieval round~\cite{zhang2023repocoder}. DraCo follows dataflow to pull in the definitions a completion will use~\cite{cheng2024draco}, and GraphCoder retrieves over a code context graph~\cite{liu2024graphcoder}. RLCoder trains a retriever with reinforcement learning~\cite{wang2024rlcoder}, while RLPG and CoCoMIC learn prompt-proposal or import structure~\cite{shrivastava2023rlpg,ding2022cocomic}. Fusion methods such as RepoFuse and CodeRAG merge several sources into one prompt~\cite{liang2024repofuse,coderag2025}. The cost profile is uniform across this line: each method spends its entire budget \emph{before} decoding, assembles one $C$, commits to it for every task, and decodes once. Even RepoCoder's iterations only refine retrieval, and fusion still merges its sources into a single prompt.

A second line decides \emph{whether} to retrieve at all. Repoformer trains a gate that skips retrieval when the model is already confident~\cite{wu2024repoformer}, and Self-RAG, FLARE, SKR, and Adaptive-RAG learn similar policies for when or how much to retrieve~\cite{asai2024selfrag,jiang2023flare,wang2023skr,jeong2024adaptiverag}. These trained gates decide retrieval timing. We instead decide \emph{which} retrieved context to trust, training-free and post hoc: the decision uses the samples a context produces, not a learned predictor of the context.

\subsection{Test-Time Scaling}
\label{sec:bg-tts}

\noindent Test-time scaling (TTS) spends budget after the prompt is fixed: sample $N$ candidates, then select one. Self-consistency draws several samples and selects by majority voting~\cite{wang2023selfconsistency}; we refer to this baseline as self-consistency@10. Repeated sampling scales coverage with $N$~\cite{brown2024monkeys,snell2024scaling}. For code, candidates are selected through generated tests, execution-based clustering, or minimum-Bayes-risk decoding~\cite{chen2023codet,li2022alphacode,shi2022mbrexec}, through trained verifiers~\cite{cobbe2021verifiers,ni2023lever}, or through Coder-Reviewer reverse reranking that scores a candidate by how well it regenerates the prompt~\cite{zhang2023coderreviewer}. These selectors are domain-general and blind to the retrieval context. They are also expensive: sampling costs about $N$ times one generation, and a reranker adds a further pass per candidate. We are aware of no instantiation for repository-level completion that exploits the retrieval context during selection.

TTS selection does not, however, depend on retrieval to work. We tested this on zero-shot prompts with no retrieved context: the prompt holds only the in-file left context, and the selector grades candidates against the in-file right context $X_r$. Sampling plus selection beats both greedy decoding and majority voting. On RepoEval-Line EM rises from 27.62 (greedy) to 32.25 (selection), on RepoEval-API from 25.33 to 27.66, against an oracle@10 of 40.31 and 34.35. The API case isolates the gain: majority voting there (24.60) falls \emph{below} greedy, yet the reverse-context verification (RCV) selector (Section~\ref{sec:approach}) lifts the same pool above greedy. The improvement comes from the selector rather than from sampling luck or vote count.

What \method{} adds to this line is not sampling or reranking as such but three specific rules and their combination: agreement among a model's own samples is used as a \emph{routing} signal across retrieval contexts rather than only as a selection signal; the code after the cursor, which every completion instance ships but causal prompting ignores, is turned into a training-free verifier by scoring the reverse likelihood with the same small model; and a fast path plus distinct-only scoring keep the whole pipeline near one generation's token budget. Self-consistency~\cite{wang2023selfconsistency} selects but does not route; best-of-$N$ with a trained verifier~\cite{cobbe2021verifiers} and forward-likelihood reranking rank candidates by how plausible they are, not by whether the file still reads coherently after them; coding agents can look up definitions and revise, but at a per-request latency far from the interactive budget of inline completion. The trade-off is concrete: accuracy rises with modest extra cost as candidates are sampled and selected, while spending more does not always help. RepoCoder's iterative retrieval sits near $2\times$ the budget without a matching gain, and exhaustive scoring of every sample adds little over scoring distinct candidates only. The open question is whether RAG and TTS can be combined so that the accuracy gains of TTS survive at near single-pass cost.

\subsection{Cost Model}
\label{sec:costmodel}

\noindent To compare methods that spend different budgets, we report \emph{relative cost}: the total LLM forward compute per task (prompt encoding, decoding, and scoring tokens) divided by that of one greedy retrieval-augmented generation, written as a multiplier. Retrieval is excluded; every retrieval-augmented method pays it once per task, independent of $N$.

The intuition is simple. The expensive part of a retrieval-augmented generation is encoding the long retrieved prompt. Prefix caching encodes it once and reuses it, so the $N$ short continuations we sample add only a little on top of one greedy generation rather than $N$ times its cost. Scoring is cheaper still: each reverse-context scoring pass is a single forward pass over a truncated context, a small fraction of one generation. A sample-then-select method therefore decomposes into a cheap sampling term and an even cheaper scoring term. Relative cost is the average number of base contexts sampled per task times $c_{\text{samp}}$, the cost of one $N$-sample batch from a context, plus the average number of distinct candidates scored times $c_{\text{rcv}}$, the cost of one reverse-context scoring pass.

Both unit costs are small in practice. Drawing a full $N$-sample batch costs only about $1.1$ times one greedy generation, because the long retrieved prefix is encoded once and the $N$ short continuations ($\le$64 tokens each) decode from the cached KV: $c_{\text{samp}} \approx 1.1$ rather than $N\times$. One reverse-context scoring pass, a single teacher-forced forward pass over a truncated context, costs only about $0.054$ of a generation: $c_{\text{rcv}} \approx 0.054$. Section~\ref{sec:approach} instantiates this accounting for \method{}, and Section~\ref{sec:results} reports measured token-level and wall-clock multipliers. The model assumes the shared prefix is cached across the $N$ samples; without caching, each sample re-encodes the prompt and the realized cost is higher (Section~\ref{sec:discussion}).

This accounting also sets the bar the method must clear. Across the self-consistency and reranking baselines above, reported test-time overhead reaches up to $1.9\times$ one generation: drawing many samples and scoring every one is what makes TTS expensive. Keeping the gain while staying near single-pass cost is the design target of \method{}.

\subsection{Motivating Study}\label{sec:bg-motivation}

\begin{figure*}[!h]
  \centering
  \includegraphics[width=0.80\textwidth]{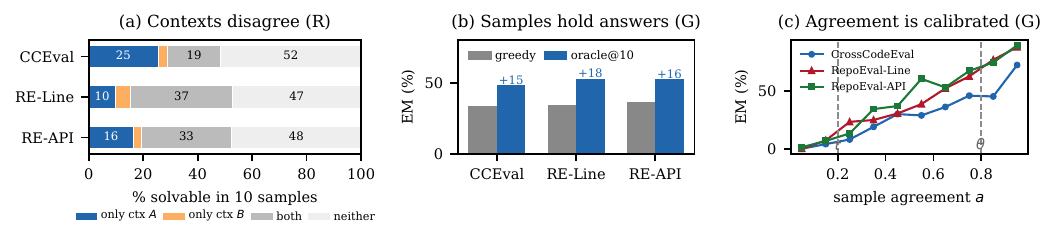}
  \caption{Motivating study with DeepSeek-Coder-1.3B on the three benchmarks, ordered by retrieval (R), generation (G), and selection. (a)~Contexts disagree: share of tasks solved within 10 samples under only the stronger context $A$, only the weaker context $B$, both, or neither; the two families cover disjoint task sets, and a per-task oracle over them gains up to $+6.0$~EM. (b)~Samples hold the answer: an oracle over 10 samples beats greedy decoding by up to $+18.4$~EM, with one or two base contexts. (c)~Agreement is calibrated: EM of the modal candidate by pool agreement $a$; dashed lines mark the routing gate $\tau$ and the skip gate $\theta$.}
  \label{fig:prelim}
\end{figure*}

\noindent We studied how repeated sampling and candidate selection behave on repository-level completion with DeepSeek-Coder-1.3B~\cite{guo2024deepseekcoder} on the three benchmarks (Section~\ref{sec:bg-task}), under the setup of Section~\ref{sec:setup}. For each benchmark we built two base contexts from different retrieval families: the stronger family on that benchmark, $A$, and the other, $B$. From each we drew $N$ samples (temperature $0.8$, top-$p$ $0.95$~\cite{holtzman2020nucleus}) plus one greedy completion. A pool \emph{solves} a task if some candidate exactly matches the reference. Fig.~\ref{fig:prelim} reports three observations on retrieval (R), generation (G), and selection.

\textbf{Observation 1 (R): No single context retrieval strategy always wins.} The two families solve substantially disjoint task sets. Partitioning tasks by which pool solves them (Fig.~\ref{fig:prelim}(a)), the stronger context $A$ uniquely solves a sizable share, yet the aggregate-weaker context $B$ still uniquely solves tasks $A$ misses. Which family is stronger flips across benchmarks, so no fixed choice transfers. An oracle picking the better context \emph{per task} gains up to $+6.0$~EM over the best fixed context. Fusion methods that merge several retrievers into one prompt~\cite{liang2024repofuse,coderag2025} dilute this per-task complementarity, which instead calls for a router that chooses a context per task.

\textbf{Observation 2 (G): The generation phase has headroom for optimization.} An oracle over 10 samples beats greedy decoding by up to $+18.4$~EM (Fig.~\ref{fig:prelim}(b)); drawing from both contexts' pools widens the gap further. On RepoEval-Line this headroom is several times the entire benefit of retrieval augmentation over in-file prompting. Though well documented for reasoning and program synthesis~\cite{wang2023selfconsistency,brown2024monkeys,chen2021codex,li2022alphacode}, it is left unused by repository-level pipelines that decode once. Sampling after retrieval exposes the answer; the open problem is picking it without an oracle.

\textbf{Observation 3 (G): Majority voting is a weak strategy, a new selector is required.} Let the \emph{agreement} $a$ of a pool be the fraction of its $N$ samples equal to the modal candidate after normalization. Agreement predicts correctness sharply (Fig.~\ref{fig:prelim}(c)): when $a\ge 0.9$ the modal candidate is correct on 72--89\% of tasks, but when $a\le 0.2$ on only 2--8\%. One free statistic thus diagnoses both a failing context (very low $a$) and a settled task (very high $a$). Calibration does not make the modal candidate a good answer, though. Vote frequency alone recovers little of the headroom and even falls below greedy on CrossCodeEval, because several distinct yet plausible candidates split the vote and the most common form is often wrong~\cite{wang2023selfconsistency,chen2023codet,zhang2023coderreviewer}. Selection needs evidence stronger than the count: the still-unused right context is a free verifier, since a correct completion should make the following code more probable.

Together, these three observations motivate \method{}'s pipeline (Section~\ref{sec:approach}): a router that picks a context per task, sampling that surfaces the answer, and a selector that reads the right context to choose it.

\section{Methodology}\label{sec:approach}

\subsection{Overview}\label{sec:approach-overview}

\begin{figure*}[t]
  \centering
  \includegraphics[width=0.95\textwidth]{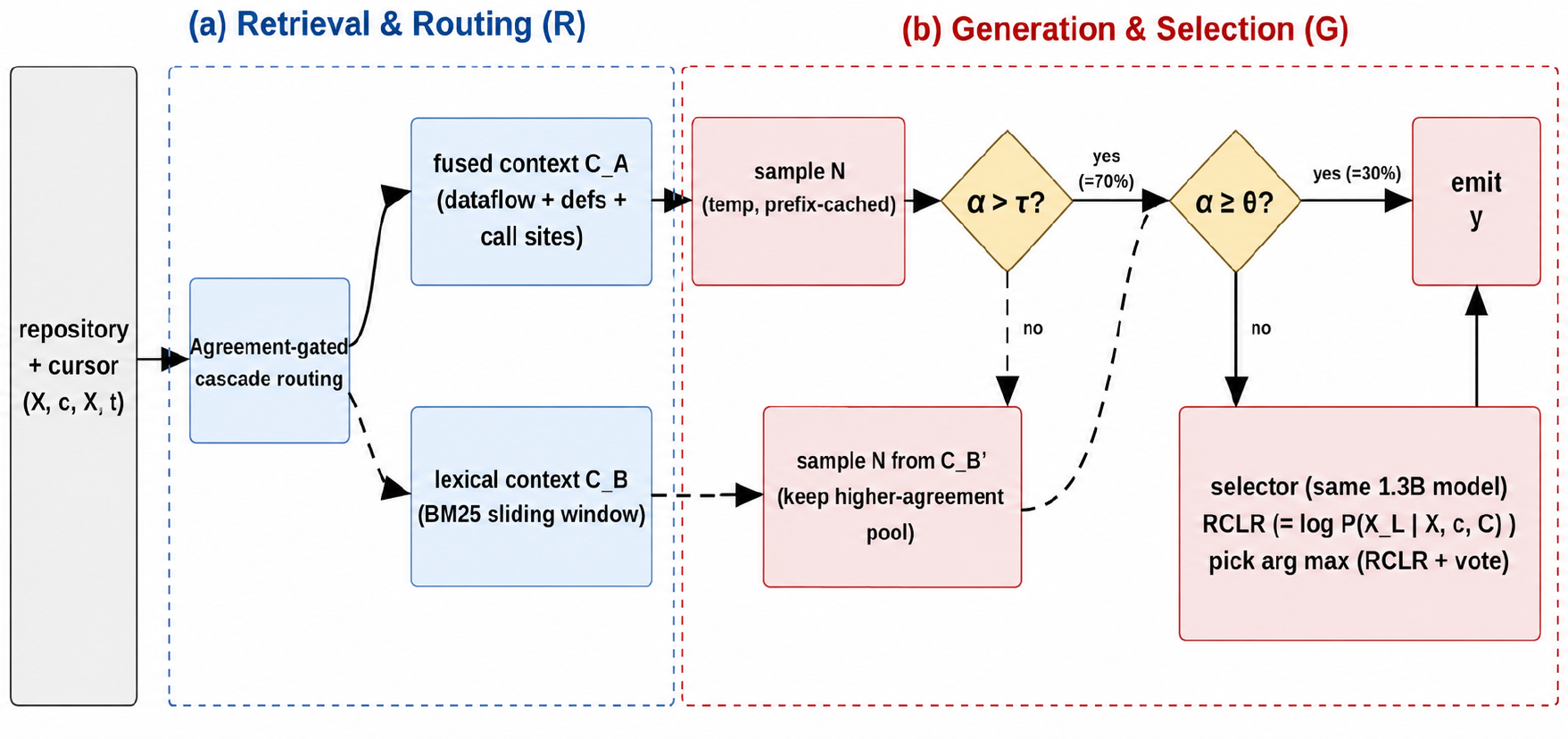}
  \caption{Overview of \method{}, in two stages. \textbf{(a)~Retrieval and routing (R)}: assemble two base contexts and route to whichever pool the samples agree on more, consulting the alternative only when the primary's agreement $a<\tau$. \textbf{(b)~Generation (G)} comprises sampling and selection: cache-aware sampling encodes the long prefix once and decodes the short continuations, and the RCV selector picks the final line---emit the modal candidate if agreement reaches $\theta$, otherwise score the distinct candidates by their reverse-context likelihood over truncated contexts and return the vote-blended argmax. Dashed paths are cost-saving shortcuts; the same small model samples and scores.}
  \label{fig:overview}
\end{figure*}

\noindent \method{} runs one training-free pass over the two stages of the pipeline, laid out left to right in Fig.~\ref{fig:overview}, with small code models only and no training or labels: the generator samples the candidates, and a fixed 1.3B scorer (the generator itself in our primary setting) verifies them. In \textbf{retrieval and routing (R)}, given the repository and the cursor, \method{} assembles two base contexts from complementary families and treats one as primary; it samples $N$ candidates from the primary, reads their agreement, and consults the alternative only when agreement is low, routing to whichever pool the samples agree on more (Fig.~\ref{fig:overview}(a), Section~\ref{sec:approach-routing}). In \textbf{generation (G)}, each consulted context is sampled in one cache-aware batch, the long prefix encoded once, the short continuations decoded from the cached state, and on the routed pool \method{} emits the modal candidate when agreement is high enough, otherwise scoring the distinct candidates by having the same model read the code after the cursor and blending that reverse-context score with vote frequency (Fig.~\ref{fig:overview}(b), Section~\ref{sec:approach-gen}).

The dashed paths in Fig.~\ref{fig:overview} are cost-saving shortcuts: consult the alternative only on disagreement, skip scoring at high agreement, and score only distinct candidates over truncated contexts. All hyperparameters are global across benchmarks and models (Section~\ref{sec:discussion}); the only per-benchmark choice is which context is primary. Cost follows the model of Section~\ref{sec:costmodel}.

Take one CrossCodeEval task as an example: its modal sample is \texttt{get\_by\_id(1)} but the correct line is \texttt{get\_by\_id(order1.id)}. The primary pool agrees at $a{=}0.6$: above $\tau$, so routing keeps the primary context, but below $\theta$, so \method{} does not trust the vote and scores the distinct candidates. Reading the code after the cursor, which references \texttt{order1.id}, the reverse-context likelihood ranks \texttt{get\_by\_id(order1.id)} above the more frequent literal-id guess, and the vote blend returns the correct line. One pass through both stages turns a wrong majority into a right answer.

\subsection{Retrieval with Agreement-Gated Cascade Routing}\label{sec:approach-routing}

\noindent \textbf{Two base retrieval strategies.}\label{sec:approach-contexts} A \emph{base context} is the cross-file evidence a retrieval strategy assembles for the prompt (Section~\ref{sec:background}). \method{} uses one structural and one lexical context, kept separate so the router can choose between them per task.

The \emph{fused context} starts from dataflow-guided retrieval following DraCo~\cite{cheng2024draco}: an import-aware def--use analysis links entities in the unfinished file to their cross-file definitions and usages, serialized into the prompt as dataflow slices. Two lightweight signals layer on this base. \emph{Identifier-definition retrieval} runs an exact-match search (\texttt{grep}) over the repository for definitions and usages of cursor-adjacent identifiers the dataflow analysis leaves unresolved, including those in the right context when available. \emph{Call-site frequency hints} annotate retrieved definitions with how often each API is invoked across the repository, surfacing representative call patterns. The recipe is base-agnostic: for API-invocation tasks a variant layers the same signals over a BM25 base, yielding a single fused-lexical context, distinct from the Concat baseline that concatenates the two families (Section~\ref{sec:setup}).

The \emph{lexical context} is BM25~\cite{zhang2023repocoder} sliding-window retrieval in the style of RepoCoder's retrieval-and-generation baseline: repository files are split into overlapping fixed-size windows, the code before the cursor is the query, and the top windows go into the prompt. As Observation~1 found, the structural and lexical families solve largely \emph{disjoint} task sets: each solves cases the other misses, and a per-task oracle over the two gains up to $+6.0$~EM over either alone. Merging them into one prompt~\cite{liang2024repofuse,coderag2025} averages away the very per-task signal routing exploits, so \method{} keeps them separate and chooses between them per task.

\noindent \textbf{Agreement-gated cascade routing.} \method{} draws $N$ samples from the primary base context and canonicalizes each to its first non-comment line. Over the resulting distinct candidates $\mathcal{D}=\{c_1,\dots,c_m\}$, the \emph{vote frequency} $v_i$ of candidate $c_i$ is its share of the $N$ samples, and the pool's \emph{agreement} is the modal share $a=\max_i v_i \in [1/N,\,1]$. When primary agreement is low, more samples from the same context rarely help: at $a\le 0.2$ the modal candidate is almost never correct (Observation~3), typically because the needed symbol or usage is absent from the prompt. \method{} therefore spends the next unit of budget on a \emph{different} context.

If the primary pool's agreement $a_1 \ge \tau$, the cascade commits to the primary. Otherwise it draws $N$ samples from the alternative base context and routes to whichever pool agrees more. The comparison is symmetric rather than a blind fallback: the primary still wins if the alternative scatters more. The gate $\tau$ keeps the alternative cheap, since the cascade falls through on a small fraction of tasks, so the expected number of sampled contexts per task stays close to one (Section~\ref{sec:costmodel}).

The cascade is asymmetric, so the primary context is \method{}'s single per-benchmark choice. The choice is automatic: the order maximizing EM on a held-out 10\% development split recovers the optimal order in every seed--benchmark combination we tried. The threshold $\tau$ doubles as insurance against a mischosen primary: a higher $\tau$ consults the alternative more often and recovers most of the lost accuracy (Section~\ref{sec:discussion}). Unlike adaptive-retrieval methods that \emph{train} a when-to-retrieve policy~\cite{wu2024repoformer,jeong2024adaptiverag,asai2024selfrag}, cascade routing is training-free: it reads a posterior signal of the generator to decide which context to trust. The cascade extends to an ordered list of $K$ contexts by advancing while agreement stays below $\tau$; we discuss $K$ in Section~\ref{sec:discussion}.

\subsection{Generation with RCV Selector}\label{sec:approach-gen}

\noindent\textbf{Cache-aware sampling.} Sampling is the step Observation~2 calls for: the pools already hold the answer far more often than greedy recovers, so the budget improves accuracy only if those samples are collected cheaply. Drawing $N$ candidates is how \method{} spends test-time compute (Section~\ref{sec:background}); the difficulty is spending it without paying $N\times$ the cost of one generation. \method{} samples each consulted context in one batch under causal prompting on $C \oplus X_l$ with nucleus sampling~\cite{holtzman2020nucleus} and a short token cap (settings in Section~\ref{sec:setup}). The long retrieved prefix is encoded once and shared; the $N$ short continuations decode from the cached state, so a full batch costs a small constant over one greedy generation rather than $N\times$ (Section~\ref{sec:costmodel}). These same $N$ samples yield both the routing signal above and the candidate pool below.

\noindent\textbf{Reverse-context verification (RCV) selector.} Observation~3 shapes selection: agreement is a calibrated confidence signal, but vote frequency alone is a weak selector when distinct plausible candidates split the vote. \method{}'s RCV selector reads the code \emph{after} the cursor to break the ties, in three steps.

\textit{Fast path.} If the routed pool's agreement satisfies $a \ge \theta$, \method{} emits the modal candidate without scoring. High-agreement modal candidates are overwhelmingly correct (Observation~3), so verification compute is better spent on harder tasks.

\textit{Reverse-context likelihood.} For the remaining tasks, vote frequency is too coarse: the pool holds several distinct yet plausible candidates with near-tied votes. \method{} breaks the ties by reading ahead. Each distinct candidate $c_i$ is scored by the likelihood the \emph{same} model assigns to the code after the cursor, conditioned on the candidate being the completion:
\begin{equation}\label{eq:rclr}
r_i \;=\; \log P_{\mathcal{M}}\!\bigl(\tilde{X}_r \,\big|\, \tilde{X}_l \oplus c_i\bigr),
\end{equation}
where $\tilde{X}_l$ and $\tilde{X}_r$ are truncated views of the in-file left and right contexts. The conditioning direction is the point. Sampling already optimizes the forward likelihood $P_{\mathcal{M}}(c_i \mid X_l)$; the reverse direction asks whether the file still reads coherently \emph{after} the candidate. Calling the wrong API or binding the wrong identifier, as above, makes the downstream code using those names measurably less probable. This reverse-context likelihood is one teacher-forced forward pass per candidate, with no decoding; a short right-context window suffices, and the scorer's left context can be dropped with little loss (Section~\ref{sec:rq-ablation}). Each call costs a small fraction of one generation (Section~\ref{sec:costmodel}).

\textit{Distinct-only scoring.} A candidate's final score depends only on the pair $(r_i, v_i)$, so scoring each \emph{distinct} candidate once is exactly equivalent to scoring all $N$ samples. This lossless deduplication cuts scoring calls well below $N$; with the fast path, it keeps the expected number of scored candidates per task small (Section~\ref{sec:costmodel}).

\textit{Blending with votes.} The selector returns
\begin{equation}\label{eq:blend}
\hat{y} \;=\; \operatorname*{arg\,max}_{c_i \in \mathcal{D}} \;\Bigl[\, 0.75\,\overline{r}_i \;+\; 0.25\,\overline{v}_i \,\Bigr],
\end{equation}
where $\overline{r}_i$ and $\overline{v}_i$ denote min--max normalization of the reverse-context likelihood and vote frequency over $\mathcal{D}$. Vote frequency acts as a consensus prior, guarding against the likelihood occasionally favoring a fluent but rarely-sampled candidate. The blend outperforms either signal alone (Section~\ref{sec:rq-ablation}); like all other knobs, the weights are global (Section~\ref{sec:discussion}).

\section{Experimental Setup}\label{sec:setup}

\subsection{Research Questions}\label{sec:setup-rqs}

\noindent We ask whether improving both stages of the pipeline, retrieval and routing (R) and generation (G) covering sampling and selection, raises accuracy at near single-pass cost. We answer three questions in Section~\ref{sec:results}, each with the metric named below.

\noindent\textbf{RQ1: Does \method{} outperform state-of-the-art single-pass and test-time-scaling baselines, and does the gain transfer across the six generators?} We answer it by exact match (EM), with edit similarity secondary (Section~\ref{sec:setup-metrics}).

\noindent\textbf{RQ2: How does the cost of \method{} compare to state-of-the-art methods?} We answer it by relative, time, and resource cost (Section~\ref{sec:setup-metrics}).

\noindent\textbf{RQ3: How much do agreement-gated routing, cache-aware sampling, and the reverse-context verification (RCV) selector contribute?} We answer it by ablating each component on the reference generator.

\subsection{Benchmarks}\label{sec:setup-benchmarks}

\noindent We evaluated \method{} on the three benchmarks of Section~\ref{sec:bg-task}: CrossCodeEval (Python; 2,124 line completions)~\cite{ding2023crosscodeeval}, RepoEval-Line (1,600 line completions), and RepoEval-API (687 API-invocation completions over 8 repositories)~\cite{zhang2023repocoder}. We used the official CrossCodeEval data and scripts, and Repoformer's released splits of RepoEval~\cite{wu2024repoformer}.

\subsection{Metrics}\label{sec:setup-metrics}

\noindent Our primary metric is \emph{exact match} (EM, \%): the fraction of tasks whose first non-comment generated line exactly matches the ground-truth line~\cite{lu2021codexglue}; \emph{edit similarity} (ES, character-level Levenshtein~\cite{lu2021codexglue}) is secondary. We confirm the significance of EM gains with per-benchmark McNemar tests on paired per-task EM against self-consistency@10 and paired-bootstrap 95\% confidence intervals of $\Delta$EM (10{,}000 resamples)~\cite{arcuri2011statistical}. For cost (RQ2) we report \emph{relative cost} (total LLM forward compute per task in multiples of one greedy generation, Section~\ref{sec:costmodel}), \emph{time cost} (wall-clock), and \emph{resource cost} (peak GPU memory).

\subsection{Baselines}\label{sec:setup-baselines}

\noindent We compared against seven single-pass configurations (zero-shot in-file, five retrieval-augmented methods, and one of our own) and one test-time-scaling method:
(1) Zero-shot in-file: greedy completion from in-file context alone;
(2) BM25: single-round sparse chunk retrieval (RepoCoder's RG-1~\cite{zhang2023repocoder}), also our lexical context;
(3) RepoCoder: iterative retrieval--generation, published two-iteration setting~\cite{zhang2023repocoder};
(4) DraCo: dataflow-guided cross-file retrieval~\cite{cheng2024draco}, underlying the dataflow component of our fused context;
(5) GraphCoder: retrieval over code context graphs~\cite{liu2024graphcoder};
(6) CodeRAG: multi-path retrieval fusion~\cite{coderag2025};
(7) Concatenated contexts (Concat): both context families in one prompt, \ie the fused context plus the lexical context's BM25 windows. This natural ``use both contexts at once'' alternative to routing is the greedy reference point for our headline gains.
(8) Self-consistency@10: majority vote over $N{=}10$ samples from the concatenated-context configuration~\cite{wang2023selfconsistency}, the strongest test-time-scaling baseline at our sampling budget.

For fair comparisons, we reproduced all baselines under identical conditions and the same input prompt-token budget, with identical decoding limits and post-processing. Numbers can therefore differ from the original papers (\eg our RepoCoder reproduction differs from RLCoder's~\cite{wang2024rlcoder}); all comparisons here are internal and mutually consistent (Section~\ref{sec:discussion}). GraphCoder's released pipeline and graph database target its own RepoEval-Updated variant, built from different repositories and task constructions. We reproduced its code-context-graph retrieval on CrossCodeEval, where it trails stronger baselines by over 16 EM, but did not port its graph database to our RepoEval splits, so GraphCoder appears only on CrossCodeEval.

\subsection{Environment}\label{sec:setup-impl}

\noindent \textit{Hardware.} All experiments ran on one machine: dual Intel Xeon Silver 4509Y CPUs (32 logical cores), 503\,GB system RAM, and a single NVIDIA H100 NVL GPU with 96\,GB memory (driver 590.48). Models ran under HuggingFace Transformers in bfloat16 with prefix caching. Time cost (wall-clock) and resource cost (peak GPU memory) in Section~\ref{sec:rq-cost} are measured on this machine; relative-cost multipliers are computed from per-task counts of consulted base contexts and scored distinct candidates under the cost model of Section~\ref{sec:costmodel}.

\textit{LLMs.} We used six opensource LLMs from four families (1.1B--7B): DeepSeek-Coder-1.3B-base (primary) and DeepSeek-Coder-6.7B-base~\cite{guo2024deepseekcoder}, SantaCoder-1.1B~\cite{allal2023santacoder}, StarCoder2-3B~\cite{lozhkov2024starcoder2}, StarCoder-7B~\cite{li2023starcoder}, and CodeLlama-7B~\cite{roziere2023codellama}. All are prompted left-to-right without fill-in-the-middle~\cite{bavarian2022fim}, so all methods consume identical prompts. SantaCoder-1.1B and StarCoder-7B, with 2048- and 8192-token context windows, are decoded at batch 1 with prompts truncated to their position limit.

\textit{Hyperparameters.} One global configuration serves all benchmarks and all six models, with no per-benchmark or per-model tuning. Sampling draws ten candidates ($N{=}10$) per consulted base context at temperature $0.8$ and nucleus $p{=}0.95$~\cite{holtzman2020nucleus}, capped at 64 new tokens, keeping the first non-comment line; greedy baselines share the prompts and the length limit. A single routing threshold, a single selection-skip threshold, and fixed blend weights complete the recipe. The only per-benchmark decision is which member of each context pair the cascade consults first: CrossCodeEval starts from the fused context, RepoEval-Line from the lexical context, RepoEval-API from the fused-lexical context, with the other family as the alternative (the pure DraCo dataflow context for RepoEval-API). We picked this one bit on a held-out 10\% development split of at least 30 tasks; across three random seeds per benchmark, all nine selections matched the full-benchmark optimum, so the split informs only this one bit. We report full task sets for comparability. The hyperparameter values appear in Section~\ref{sec:discussion}, and the routing sensitivity analysis in Section~\ref{sec:rq-ablation}.

\textit{Selector.} The RCV selector always uses DeepSeek-Coder-1.3B, even for larger generators, so \method{} stays training-free and keeps the selection cost at a reasonable level. Full inference configurations, prompt budgets, and retrieval parameters are in the artifact.

\section{Evaluation}\label{sec:results}

\noindent Improving both stages, retrieval and routing (R) and generation (G, which spans sampling and selection), raises repository-level completion accuracy at a small, bounded inference budget, and the gain holds across generators. \method{} pairs agreement-gated cascade routing on the retrieval side and cache-aware sampling on the generation side with the reverse-context verification (RCV) selector. We answer the three research questions of Section~\ref{sec:setup-rqs} in turn: accuracy and generalization (Section~\ref{sec:results-generalize}), cost (Section~\ref{sec:rq-cost}), and component attribution (Section~\ref{sec:rq-ablation}). Unless noted, the reference generator is DeepSeek-Coder-1.3B~\cite{guo2024deepseekcoder} and the metric is exact match (EM; Section~\ref{sec:setup-metrics}). The same DeepSeek-Coder-1.3B selector serves every generator without retraining. All baselines are self-reproduced under identical prompting, context budgets, and evaluation (Section~\ref{sec:setup}).

\begin{table}[t]
\caption{\textbf{RQ1}: cross-model exact match (EM, \%) of \method{} against retrieval and test-time-scaling baselines on all six generators. Concat, concatenated-context greedy (both context families in one prompt); SC@10, self-consistency@10. Per model column, the \colorbox{g1}{\textbf{best}}, \colorbox{g2}{second}, and \colorbox{g3}{third} non-oracle scores are shaded. Oracle@10 is the best-of-ten upper bound.}
\label{tab:crossmodel}
\centering\setlength{\tabcolsep}{3pt}\renewcommand{\arraystretch}{1.0}
\resizebox{\columnwidth}{!}{%
\begin{tabular}{l cccccc}
\toprule
\multicolumn{7}{c}{\textit{CrossCodeEval}}\\
\midrule
Method & DS-1.3B & DS-6.7B & SC-7B & CL-7B & SC2-3B & Santa \\
\midrule
\textit{Oracle@10} & \textit{44.73} & \textit{50.75} & \textit{47.41} & \textit{46.00} & \textit{45.48} & \textit{30.89} \\
Zero-shot & 6.12 & 9.46 & 6.97 & 7.30 & 6.64 & 3.81 \\
BM25 & 14.55 & 17.98 & 15.49 & 15.30 & 15.58 & 9.32 \\
RepoCoder & 17.47 & 21.52 & 18.69 & 19.16 & 19.44 & 12.81 \\
CodeRAG & 31.64 & \cellcolor{g2}39.17 & \cellcolor{g3}32.20 & \cellcolor{g2}34.37 & \cellcolor{g2}32.11 & 14.17 \\
DraCo & 30.56 & \cellcolor{g3}38.23 & 30.32 & \cellcolor{g3}33.57 & 30.65 & 12.95 \\
GraphCoder & 14.17 & 18.50 & 15.44 & 15.68 & 15.91 & 10.50 \\
Concat & \cellcolor{g2}33.47 & 34.09 & 30.32 & 29.90 & 28.01 & \cellcolor{g3}17.28 \\
SC@10 & \cellcolor{g3}31.97 & 38.09 & \cellcolor{g2}33.85 & 33.47 & \cellcolor{g3}31.36 & \cellcolor{g2}20.20 \\
Ours & \cellcolor{g1}\textbf{33.99} & \cellcolor{g1}\textbf{39.55} & \cellcolor{g1}\textbf{35.36} & \cellcolor{g1}\textbf{35.26} & \cellcolor{g1}\textbf{33.85} & \cellcolor{g1}\textbf{22.55} \\
\bottomrule\\[-6pt]
\toprule
\multicolumn{7}{c}{\textit{RepoEval-Line}}\\
\midrule
Method & DS-1.3B & DS-6.7B & SC-7B & CL-7B & SC2-3B & Santa \\
\midrule
\textit{Oracle@10} & \textit{47.00} & \textit{51.69} & \textit{48.06} & \textit{47.12} & \textit{45.81} & \textit{40.19} \\
Zero-shot & 28.00 & 31.25 & 28.56 & 23.69 & 25.25 & 24.38 \\
BM25 & 32.25 & 33.12 & 31.94 & 31.62 & 28.00 & 27.38 \\
RepoCoder & \cellcolor{g3}34.56 & \cellcolor{g3}34.44 & \cellcolor{g3}33.88 & \cellcolor{g2}33.62 & \cellcolor{g3}30.31 & \cellcolor{g3}28.81 \\
CodeRAG & 31.37 & 33.94 & 31.00 & 31.13 & 29.19 & 25.62 \\
DraCo & 29.94 & \cellcolor{g3}34.44 & 30.38 & 29.12 & 27.56 & 25.50 \\
Concat & 34.31 & 27.00 & 27.00 & 24.00 & 24.94 & 24.25 \\
SC@10 & \cellcolor{g2}36.06 & \cellcolor{g2}35.12 & \cellcolor{g2}34.69 & \cellcolor{g3}32.94 & \cellcolor{g2}32.12 & \cellcolor{g2}30.00 \\
Ours & \cellcolor{g1}\textbf{41.75} & \cellcolor{g1}\textbf{43.88} & \cellcolor{g1}\textbf{41.44} & \cellcolor{g1}\textbf{40.50} & \cellcolor{g1}\textbf{39.88} & \cellcolor{g1}\textbf{36.06} \\
\bottomrule\\[-6pt]
\toprule
\multicolumn{7}{c}{\textit{RepoEval-API}}\\
\midrule
Method & DS-1.3B & DS-6.7B & SC-7B & CL-7B & SC2-3B & Santa \\
\midrule
\textit{Oracle@10} & \textit{49.34} & \textit{52.11} & \textit{50.95} & \textit{46.87} & \textit{45.41} & \textit{37.12} \\
Zero-shot & 25.33 & 27.95 & 25.04 & 21.11 & 20.67 & 19.94 \\
BM25 & 35.37 & \cellcolor{g3}40.17 & \cellcolor{g3}33.04 & \cellcolor{g3}33.04 & \cellcolor{g3}30.86 & 23.58 \\
RepoCoder & 32.61 & 33.33 & 32.02 & 31.73 & 28.24 & \cellcolor{g3}24.02 \\
CodeRAG & 34.35 & 40.03 & 29.99 & 31.15 & 29.84 & 22.42 \\
DraCo & 25.47 & 29.84 & 25.76 & 23.44 & 22.42 & 19.94 \\
Concat & \cellcolor{g3}36.54 & 31.00 & 27.66 & 25.76 & 24.45 & 21.11 \\
SC@10 & \cellcolor{g2}37.99 & \cellcolor{g2}40.32 & \cellcolor{g2}38.86 & \cellcolor{g2}33.48 & \cellcolor{g2}33.62 & \cellcolor{g2}26.20 \\
Ours & \cellcolor{g1}\textbf{43.52} & \cellcolor{g1}\textbf{47.74} & \cellcolor{g1}\textbf{44.69} & \cellcolor{g1}\textbf{42.36} & \cellcolor{g1}\textbf{39.59} & \cellcolor{g1}\textbf{32.90} \\
\bottomrule
\end{tabular}}
\end{table}

\subsection{RQ1: Does \method{} outperform state-of-the-art single-pass and test-time-scaling baselines, and does the gain transfer across the six generators?}\label{sec:results-generalize}

\noindent \method{} is the best non-oracle method in every column of Table~\ref{tab:crossmodel}: it improves all 18 model--benchmark combinations over greedy decoding and self-consistency@10~\cite{wang2023selfconsistency}. On all six generators it tops every retrieval baseline (BM25, RepoCoder, CodeRAG, DraCo, and GraphCoder) in every benchmark. The six generators span four model families and 1.1B--7B parameters~\cite{guo2024deepseekcoder,li2023starcoder,lozhkov2024starcoder2,roziere2023codellama,allal2023santacoder}, yet every run uses the same global hyperparameters ($N{=}10$, $\tau{=}0.2$, $\theta{=}0.8$, fixed blend weights) and one fixed DeepSeek-Coder-1.3B selector. Averaged over the six models, the gain over greedy is +4.6 EM on CrossCodeEval, +13.7 on RepoEval-Line, and +14.0 on RepoEval-API: +10.8 EM across all 18 cells. Over self-consistency@10 the mean gain is +5.3 EM overall, +1.9/+7.1/+6.7 per benchmark. One small selector handles generators five times its size without retuning, because verifying a candidate against the right context is easier than generating it, so no per-generator verifier or training is needed.

The 1.3B selector with sampling and selection can outrun a much larger generator's single pass. On RepoEval-Line and RepoEval-API the 1.3B reaches 41.75 and 43.52 EM, above DeepSeek-Coder-6.7B's best single-pass baseline (34.44 and 40.17) and above the 6.7B self-consistency@10 vote: a 1.3B generator at roughly 1.5$\times$ compute beats a model five times larger run once.

\noindent \textbf{Reference-model standing.} The DeepSeek-Coder-1.3B column of Table~\ref{tab:crossmodel} places \method{} among retrieval baselines on the reference model. Zero-shot completion collapses on CrossCodeEval (6.12 EM): cross-file context, rather than model capacity, is the bottleneck~\cite{ding2023crosscodeeval}. No prior retrieval strategy wins everywhere: CodeRAG~\cite{coderag2025} leads CrossCodeEval among baselines (31.64) yet trails plain BM25 on RepoEval-Line (31.37 vs.\ 32.25), and DraCo~\cite{cheng2024draco} (30.56) barely clears zero-shot on RepoEval-API (25.47 vs.\ 25.33). \method{} tops every retrieval baseline on all three benchmarks (33.99 / 41.75 / 43.52 EM) and strictly dominates RepoCoder's iterative two-pass pipeline~\cite{zhang2023repocoder} on each. ES follows the same ordering (76.08 on CrossCodeEval, 71.13 on RepoEval-Line, 74.09 on RepoEval-API).

\method{} also beats the budget-matched test-time-scaling baseline. Self-consistency@10 votes over ten samples but improves greedy only modestly on RepoEval-Line (+1.75) and RepoEval-API (+1.45), and \emph{degrades} CrossCodeEval (31.97 vs.\ 33.47): votes alone fail where candidate pools hold many distinct-but-plausible lines. \method{} beats it by +2.0 / +5.7 / +5.5 EM, on the same ten-sample budget plus selector scoring.

\noindent \textbf{Significance and the CrossCodeEval boundary.} The shared ten-sample budget admits paired tests~\cite{arcuri2011statistical} against the primary-context self-consistency baseline, the majority vote over the ten primary-context samples (31.54/34.31/36.97 EM on CrossCodeEval/RepoEval-Line/RepoEval-API). McNemar gives $p=1.1\times10^{-4}$ on CrossCodeEval (113 tasks solved only by \method{} vs.\ 61), $p=2.9\times10^{-23}$ on RepoEval-Line (130 vs.\ 11), and $p=1.0\times10^{-8}$ on RepoEval-API (52 vs.\ 7). Paired-bootstrap 95\% CIs of $\Delta$EM (10{,}000 resamples) are $[+1.27,+3.67]$, $[+6.06,+8.88]$, and $[+4.51,+8.73]$, all excluding zero, and all three $p$-values survive Bonferroni correction at $\alpha{=}0.05/3$.

The CrossCodeEval gain on the reference model is the smallest in Table~\ref{tab:crossmodel}'s DeepSeek-Coder-1.3B column, but it is a property of that one generator rather than the benchmark: DeepSeek-Coder-1.3B has an unusually strong CrossCodeEval greedy baseline. Across the other five generators the CrossCodeEval gain over greedy averages +5.4 EM, range +5.0 to +5.8 (Table~\ref{tab:crossmodel}). The benchmark is not resistant to test-time scaling; the reference model is simply close to its single-pass ceiling there.

Other test-time-scaling work reports best-of-$N$ or pass@$k$ as an upper bound that a real selector cannot reach~\cite{brown2024monkeys,snell2024scaling}. Following that convention, Oracle@10 is the top row of each block in Table~\ref{tab:crossmodel}. The oracle reaches 44.73 / 47.00 / 49.34 EM on the reference model, well above \method{}, so selection is far from saturated. Section~\ref{sec:rq-ablation} dissects which component captures how much of this headroom.

\subsection{RQ2: How does the cost of \method{} compare to state-of-the-art methods?}\label{sec:rq-cost}

\begin{table}[t]
\caption{\textbf{RQ2}: per-method inference cost on each generator (means over the three benchmarks; 40 prompts/benchmark, one H100). Single-pass methods and RepoCoder decode one completion at a time, so their peak GPU memory matches greedy decoding; SC@10 and \method{} hold the same ten-sample batch; the shorter RCV scoring pass does not increase peak memory, so peak GPU memory is identical to SC@10.}
\label{tab:cost}
\centering\scriptsize\setlength{\tabcolsep}{3.5pt}\renewcommand{\arraystretch}{1.05}
\begin{tabular}{l cccccc}
\toprule
Method & DS-1.3B & DS-6.7B & SC-7B & CL-7B & SC2-3B & Santa \\
\midrule
\multicolumn{7}{c}{\textit{Execution time} (normalized to one greedy generation)} \\
\midrule
Single-pass RAG & 1.00 & 1.00 & 1.00 & 1.00 & 1.00 & 1.00 \\
RepoCoder & ${\sim}2.0$ & ${\sim}2.0$ & ${\sim}2.0$ & ${\sim}2.0$ & ${\sim}2.0$ & ${\sim}2.0$ \\
SC@10 & 1.29 & 2.22 & 2.04 & 2.27 & 1.76 & 1.21 \\
Ours & 1.45 & 2.49 & 2.30 & 2.54 & 1.97 & 1.38 \\
\midrule
\multicolumn{7}{c}{\textit{Peak GPU memory} (GB)} \\
\midrule
Single-pass RAG & 3.7 & 16.1 & 15.3 & 16.1 & 6.5 & 2.5 \\
RepoCoder & 3.7 & 16.1 & 15.3 & 16.1 & 6.5 & 2.5 \\
SC@10 & 12.8 & 39.1 & 20.3 & 39.1 & 10.6 & 4.6 \\
Ours & 12.8 & 39.1 & 20.3 & 39.1 & 10.6 & 4.6 \\
\bottomrule
\end{tabular}
\end{table}

\noindent \method{} spends a small, bounded budget that the cost model of Section~\ref{sec:costmodel} predicts. The \emph{token-level compute} multiplier is ${\sim}1.45\times$ one greedy generation on every generator: the cascade touches only 1.09--1.12 base contexts per task, so sampling costs ${\approx}1.2\times$; the high-agreement skip fires on 24.1--34.6\% of tasks; and distinct-only scoring prices the rest at $0.054\times$ per candidate. Disabling all three savings rules raises this to ${\sim}1.9\times$ for at most $+0.73$~EM (Section~\ref{sec:rq-ablation}), so the gated pipeline is near-lossless and stays below RepoCoder's iterative ${\sim}2\times$.

\noindent\textbf{Cost in context} (Table~\ref{tab:cost}). Every single-pass retrieval baseline (zero-shot, BM25, CodeRAG, DraCo, GraphCoder, Concat) decodes once, so all share the $1\times$ generation-time row; they differ in what is retrieved. Retrieval is not free but it is small: on RepoEval-API (40 tasks, single CPU thread) a BM25 window query takes 0.10\,s, DraCo's dataflow retrieval over the prebuilt graph 0.12\,s, and our call-site and grep signals 0.30\,s, so the fused-lexical context \method{} starts from costs 0.51\,s per task and the alternative context, built on the 11\% of tasks that fall through, adds 0.01\,s on average (index and graph construction are per-repository and offline: 0.1--0.6\,s and minutes, respectively). End to end, one \method{} completion takes 2.2\,s on DeepSeek-Coder-1.3B and 4.5\,s on the 7B generators, against 1.2--1.5\,s for a BM25 single pass ($1.9$--$3.1\times$); retrieval is 12--42\% of \method{}'s time, most of it on the smallest models. RepoCoder's two-pass loop roughly doubles that. \method{} costs essentially what SC@10 does (the budget-matched test-time-scaling baseline), since both draw the same ten samples and distinct-only RCV scoring adds only $0.1$--$0.3\times$ compute on top, yet \method{} is $+5.3$ EM more accurate. Its compute (${\sim}1.45\times$) stays below RepoCoder's iterative ${\sim}2\times$. Peak GPU memory is identical to SC@10's: the ten-sample sampling batch dominates memory, and the shorter RCV scoring pass does not raise the peak. Wall-clock, by contrast, tracks model size rather than the compute multiplier: batched sampling is nearly free on the memory-bound small generators ($1.33$--$1.56\times$ on DeepSeek-Coder-1.3B and SantaCoder) but costs $1.9$--$2.9\times$ on the compute-bound 7B ones, where the ten samples no longer hide behind memory latency.

\noindent\textbf{Resource cost} (Table~\ref{tab:cost}) grows with the KV cache of the ten-sample batch: multi-query-attention models keep it small (StarCoder-7B $15.3{\to}20.3$~GB, SantaCoder $2.5{\to}4.6$~GB), while the multi-head models grow more (DeepSeek-Coder-6.7B and CodeLlama-7B $16.1{\to}39.1$~GB). The extra memory is transient, freed after each task, and fits one accelerator, but it is not free in a serving setting: at a fixed memory budget the ten-sample batch cuts the number of concurrent requests a multi-head 7B model can hold by about $2.4\times$ (39\,GB vs.\ 16\,GB) and a multi-query model by about $1.3\times$; a server can trade this for latency by sampling in two rounds of five, which halves the peak at ${\approx}10\%$ more time.

\noindent\textbf{Compute substitutes for parameters.} DeepSeek-Coder-1.3B with \method{} reaches 41.75~EM on RepoEval-Line and 43.52 on RepoEval-API, ahead of the $5\times$-larger DeepSeek-Coder-6.7B's best single-pass baseline (34.44/40.17) and of 6.7B self-consistency@10 (35.12/40.32), at a fraction of its memory. The trade is workload-dependent: on CrossCodeEval the larger model keeps its lead (39.17 vs.\ 33.99). A fixed inference budget spent on sampling, routing, and selection can return more EM than the same budget spent on larger weights.

\subsection{RQ3: How much do agreement-gated routing, cache-aware sampling, and the reverse-context verification (RCV) selector contribute?}\label{sec:rq-ablation}

\begin{table}[t]
\caption{\textbf{RQ3}, routing (R) (DeepSeek-Coder-1.3B, EM \%): context-choice strategies under a matched majority@10 selector. Per benchmark the \emph{primary} context is the dev-selected family---structural (fused) on CrossCodeEval, lexical (BM25) on RepoEval-Line, fused-lexical on RepoEval-API---and the \emph{alternative} is the other family; ``Ours'' routes between them by agreement. \textbf{Bold}: best per column.}
\label{tab:abl-r}
\centering\footnotesize\setlength{\tabcolsep}{4pt}
\resizebox{\columnwidth}{!}{%
\begin{tabular}{lccc}
\toprule
Context strategy & CrossCodeEval & RepoEval-Line & RepoEval-API \\
\midrule
Primary context only & 31.54 & 34.31 & \textbf{36.97} \\
Alternative context only & 14.17 & 31.94 & 25.91 \\
Ours & \textbf{31.83} & \textbf{37.06} & \textbf{36.97} \\
\bottomrule
\end{tabular}}
\end{table}

\begin{table}[t]
\caption{\textbf{RQ3}, sampling (G) (DeepSeek-Coder-1.3B, EM \%): exact match and relative compute vs.\ the sample budget $N$ on the primary context, cascade routing disabled (Section~\ref{sec:costmodel}); the full \method{} pipeline appears in Table~\ref{tab:crossmodel}. $N{=}10$ is our deployed budget.}
\label{tab:abl-g}
\centering\footnotesize\setlength{\tabcolsep}{4pt}
\resizebox{\columnwidth}{!}{%
\begin{tabular}{lcccc}
\toprule
Sample budget & CrossCodeEval & RepoEval-Line & RepoEval-API & Rel.\ cost \\
\midrule
$N{=}5$ & 30.56 & 36.12 & 38.72 & $1.21\times$ \\
$N{=}10$ (Ours) & 32.36 & 38.10 & 40.32 & $1.45\times$ \\
$N{=}15$ & 32.50 & 39.60 & 42.07 & $1.61\times$ \\
\bottomrule
\end{tabular}}
\end{table}

\begin{table}[t]
\caption{\textbf{RQ3}, selection (G) (DeepSeek-Coder-1.3B, EM \%): selectors on a fixed candidate pool. \textbf{Bold}: best per column.}
\label{tab:abl-s}
\centering\footnotesize\setlength{\tabcolsep}{4pt}
\resizebox{\columnwidth}{!}{%
\begin{tabular}{lccc}
\toprule
Selector & CrossCodeEval & RepoEval-Line & RepoEval-API \\
\midrule
Majority vote & 31.97 & 36.06 & 37.99 \\
Reverse-context likelihood & 31.54 & 36.94 & 42.65 \\
Ours & \textbf{32.82} & \textbf{38.12} & \textbf{43.67} \\
\bottomrule
\end{tabular}}
\end{table}

\begin{figure*}[t]
\centering
\includegraphics[width=0.74\textwidth]{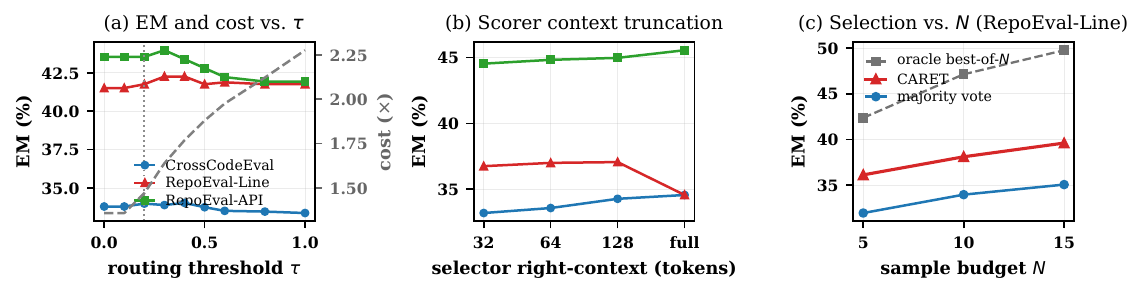}
\caption{Sensitivity of \method{} (DeepSeek-Coder-1.3B). (a)~EM and relative cost vs.\ routing threshold $\tau$ (default $\tau{=}0.2$, dashed). (b)~EM vs.\ the selector's right-context budget; 128 tokens is near-lossless and truncation helps RepoEval-Line. (c)~Selection vs.\ sample budget $N$ on RepoEval-Line: vote saturates while \method{} tracks the best-of-$N$ oracle.}
\label{fig:sensitivity}
\end{figure*}

\noindent We ablate routing, sampling, and selection in turn on the reference generator (Tables~\ref{tab:abl-r}--\ref{tab:abl-s}), each isolated under a matched protocol.

\noindent \textbf{Routing (R).} Table~\ref{tab:abl-r} holds selection at a majority vote and varies the context choice. The two base contexts solve different tasks, so agreement-gated routing beats either fixed context on RepoEval-Line (37.06 vs.\ 34.31 and 31.94). Under the full \method{} selector the marginal effect is smaller: disabling routing ($\tau{=}0$) while keeping sampling, the fast path, and RCV costs $-0.19$/$-0.25$/$0.00$~EM on the reference generator, and across the five other generators $+0.9$ to $+1.6$~EM on RepoEval-Line but $-0.6$ to $+0.4$ on the other two benchmarks (Table~\ref{tab:route-full}). The two ablations differ because RCV already repairs part of what a wrong context does to the pool: with majority voting the pool's errors go through, with RCV the alternative context matters only when its candidates are both different and verifiable. The primary order is chosen on a 10\% development split; excluding that split (three seeds) moves \method{}'s EM by $-0.8$ to $+1.7$ and its margin over SC@10 by at most $\pm0.8$, and the margin stays positive in all 45 model--benchmark--seed cases (minimum $+1.3$). Routing is a cheap safeguard: with a swapped primary, CrossCodeEval collapses to 19.87~EM at $\tau{=}0.2$ but $\tau{=}1$ recovers 32.25, and the 10\% dev split picked the primary correctly in 9/9 benchmark--seed combinations, so the cheap default is safe (Fig.~\ref{fig:sensitivity}(a)).

\begin{table}[t]
\caption{\textbf{RQ3}, routing under the \emph{full} \method{} selector (EM~\%): agreement-gated routing on vs.\ off (primary context only), everything else unchanged.}
\label{tab:route-full}
\centering\footnotesize\setlength{\tabcolsep}{3.5pt}
\resizebox{\columnwidth}{!}{%
\begin{tabular}{l cc cc cc}
\toprule
 & \multicolumn{2}{c}{CrossCodeEval} & \multicolumn{2}{c}{RepoEval-Line} & \multicolumn{2}{c}{RepoEval-API} \\
Generator & off & on & off & on & off & on \\
\midrule
DS-6.7B & 39.83 & 39.83 & 42.81 & 43.94 & 47.74 & 47.60 \\
SC-7B & 35.40 & 35.55 & 40.31 & 41.19 & 44.69 & 44.54 \\
CL-7B & 35.59 & 35.64 & 39.06 & 40.62 & 42.50 & 42.50 \\
SC2-3B & 33.85 & 34.04 & 38.31 & 39.56 & 39.74 & 39.16 \\
Santa & 22.46 & 22.83 & 35.44 & 36.06 & 32.75 & 33.04 \\
\bottomrule
\end{tabular}}
\end{table}

\noindent \textbf{Sampling (G).} Table~\ref{tab:abl-g} varies the sample budget against its relative compute. A best-of-$N$ oracle would reach 44.94/49.75/53.71~EM, far above any single decode, the headroom selection targets. We fix $N{=}10$, the budget our self-consistency baseline also uses: CrossCodeEval has saturated there ($+1.80$~EM from $N{=}5$ to $10$, then only $+0.14$ to $15$), and while RepoEval keeps gaining to $N{=}15$ ($+1.5$/$+1.8$~EM), that step costs $+0.16\times$ compute and $50\%$ more peak memory, which we judge not worth it. Removing sampling entirely reduces \method{} to the concatenated-contexts greedy baseline: $-7.44$~EM on RepoEval-Line and $-6.98$ on RepoEval-API.

\noindent \textbf{Selection (G).} Table~\ref{tab:abl-s} fixes the candidate pool and varies the selector. Reverse-context likelihood beats majority voting on every benchmark, and blending the two is best (43.67 vs.\ 37.99~EM on RepoEval-API). Majority voting alone is barely above greedy on RepoEval and \emph{below} it on CrossCodeEval: calibrated agreement localizes the hard tasks, but only reading the right context picks the answer. At $N{=}10$ the selector captures 16\%/50\%/49\% of the vote-to-oracle headroom, overriding a wrong modal candidate on 105/112/48 tasks (Fig.~\ref{fig:sensitivity}(c)). Its right context is essential but short: 128 tokens is near-lossless and even helps RepoEval-Line, while the scorer's left context can be dropped entirely, so each scoring call is one cached forward pass at $0.054\times$ a generation (Fig.~\ref{fig:sensitivity}(b)).

\noindent \textbf{Cost savings are near-lossless.} Disabling all three savings rules raises cost to ${\sim}1.9\times$ for at most $+0.73$/$+0.69$~EM on RepoEval-API/Line, while CrossCodeEval \emph{drops} $0.94$, since its high-agreement skip fires on 34.6\% of tasks and protects confident majorities. Exhaustive scoring adds no accuracy the gated ${\sim}1.45\times$ pipeline does not already reach.

\section{Threats to Validity and Discussion}
\label{sec:discussion}

\noindent The evaluation makes broad claims from one implementation and a single primary metric, so we set out its limits before drawing conclusions. We organize them as construct, internal, and external validity, then discuss the reasoning behind the global recipe. Two properties bound most of the risk: every method runs under the same prompts, budget, and decoding, and \method{} exposes no tuned knob beyond a single per-benchmark context choice, so the gains are hard to attribute to a selective harness or to hyperparameter fitting.

\subsection{Threats to Validity}

\noindent\textbf{Construct validity.}
Exact match and edit similarity~\cite{lu2021codexglue} measure surface agreement with a single reference rather than functional correctness; line and API-invocation granularity mitigates this, since a statement admits a narrow canonical form unlike function bodies~\cite{zhang2023repocoder,ding2023crosscodeeval}, and all methods share the same first-line normalization. The residual risk that EM gains may not map to practical benefit~\cite{peng2023copilot} is shared by every compared method. The relative-cost multiplier is itself a construct: it counts forward tokens and assumes prefix caching across the $N$ samples; without caching, overhead is higher. The ${\sim}1.45\times$ token multiplier is model-independent, but the realized wall-clock is not (Section~\ref{sec:rq-cost}): it falls below the multiplier on memory-bound small generators and rises above it ($1.9$--$2.9\times$) on compute-bound 7B ones, so we report both and release per-task counts to re-derive cost under any accounting.

\noindent\textbf{Internal validity.}
Reproduced numbers differ from published ones (our RepoCoder differs from RLCoder's~\cite{wang2024rlcoder} through different data processing and prompts), so we compare only within our pipeline, with identical prompt budget, decoding, and scoring for every method. \method{}'s global hyperparameters are fixed once for every model and benchmark; only the primary base context is per-benchmark, chosen on a 10\% development split that picked the correct primary in nine of nine seed--benchmark combinations, and raising $\tau$ recovers most of a mischoice (Section~\ref{sec:rq-ablation}). Benchmark repositories may overlap pretraining data, but all methods decode from the same frozen models, so memorization could inflate absolute scores, not deltas. We report one sampling run; task-set sizes (2{,}124/1{,}600/687) and paired tests~\cite{arcuri2011statistical} mitigate a lucky run: against the primary-context self-consistency baseline, McNemar rejects equal accuracy and 95\% paired-bootstrap EM-gain intervals exclude zero on all three benchmarks (Section~\ref{sec:results-generalize}).

\noindent\textbf{External validity.}
Our main study covers Python. A sanity check on CrossCodeEval's Java split~\cite{ding2023crosscodeeval} (2{,}139 line tasks, BM25 context, same hyperparameters) supports transfer of selection: sampling plus selection lifts the 1.3B reference from 16.27 EM (greedy) and 16.74 (self-consistency@10) to 19.17. Two qualifications hold: the scorer needed a 1{,}000-character left-context tail on Java, whereas Python tolerates dropping it entirely (Section~\ref{sec:rq-ablation}); and the fused context's dataflow component is Python-specific, so routing was not exercised and other languages need their own dataflow extractors. Granularity is a second limit: we studied line and API-invocation completion, while function bodies change both candidate diversity and metric validity. Generators span 1.1B--7B base models from four families; larger or instruction-tuned models may sample differently, though gains in all 18 combinations suggest the mechanisms are not model-specific. The selector assumes code after the completion point, the common case in these benchmarks, in fill-in-the-middle training~\cite{bavarian2022fim}, and in prior work~\cite{wu2024repoformer}. It degrades gracefully as that context shrinks: a 32-token window costs 1.02 EM on RepoEval-API and 1.37 on CrossCodeEval and helps RepoEval-Line (Fig.~\ref{fig:sensitivity}). With no right context, selection falls back to vote frequency on the routed pool (33.99$\to$31.83 / 41.75$\to$34.94 / 43.52$\to$37.26 EM): above greedy on RepoEval-Line and API, below on CrossCodeEval, and below full \method{} everywhere, confirming that the right context drives the selection gains.

\subsection{Discussion}

\noindent\textbf{Global hyperparameters.}
\method{} fixes one recipe and reuses it across every model and benchmark: routing threshold $\tau{=}0.2$, skip gate $\theta{=}0.8$, blend weights 0.75/0.25, sample budget $N{=}10$. The four values were set once on the reference generator (DeepSeek-Coder-1.3B) and frozen before any other model was run, so the other five generators and the closed models are held out with respect to them. The surface is flat: sweeping $\theta\in\{0.6,\dots,1.0\}$ and the RCV weight $w\in\{0.5,\dots,1.0\}$ offline on the stored candidate pools of the five non-reference generators, the default lies within $0.08$~EM of the best cell averaged over 15 model--benchmark pairs and at most $0.87$~EM below the per-pair optimum; $\theta$ mainly moves cost (scored share 0.24--0.29$\times$), not accuracy. The only per-benchmark choice is which context is primary, a binary decision made on a 10\% development split that picked the same order in all nine seed--benchmark cases; the development tasks are part of the reported test sets, and excluding them changes no comparison (\S\ref{sec:rq-ablation}). For a new repository without labelled completions the default is the fused structural context, which was primary on two of three benchmarks, and the cascade recovers most of the loss of a wrong choice (Fig.~\ref{fig:sensitivity}(a)).

\noindent\textbf{Completion granularity.}
\method{} is built for line and API-invocation completion, and its mechanisms do not transfer unchanged to function bodies. On RepoEval-Function (455 tasks, DeepSeek-Coder-1.3B, ten samples at the same temperature), 96\% of candidate pools contain no repeated candidate (9.85 distinct of 10), so agreement carries no signal for routing or for the fast path, and the sampled pool itself is worse than greedy decoding: the best of ten samples reaches 29.5 edit similarity against 48.4 for one greedy pass, and reverse-context selection over that pool (45.6) does not recover the gap. Function-level test-time scaling needs a different canonicalization (semantic rather than string agreement), a lower sampling temperature, and execution-based selection; we leave it to future work and restrict our claims to the line and API level.

\noindent\textbf{Number of base contexts and the sample budget.}
The cascade routes between $K{=}2$ base contexts, the sweet spot: a third adds only $+0.1$ EM for $+0.2\times$ compute, so the marginal context does not pay for itself. The sample budget $N$ trades accuracy for cost steeply at first and then flattens. An $N\in\{5,10,15\}$ study shows accuracy saturating around $N{=}10$ (CrossCodeEval gains $+1.80$ EM from $N{=}5$ to $10$ but only $+0.14$ to $N{=}15$; Fig.~\ref{fig:sensitivity}(c)), so we draw ten samples: enough to surface the answer and calibrate agreement, few enough to keep the budget near one greedy generation.

\section{Conclusion and Future Work}
\label{sec:conclusion}

\noindent We presented \method{}, a training-free framework that improves both stages of retrieval-augmented completion: retrieval and routing (R), and generation (G). It allocates a small inference budget across them: retrieval feeds two base contexts, sampling draws candidates that agreement routes between them, and reverse-context likelihood selects the final line. Routing and selection read two signals the completion setting provides for free: agreement among the model's own samples and the file's right context. Under one global recipe, \method{} improves all 18 model--benchmark combinations across six generators (1.1B--7B) and three benchmarks by $+10.8$ EM on average over greedy decoding and $+5.3$ over self-consistency@10, at ${\sim}1.45\times$ the token-level compute of one generation, with measured wall-clock from $1.3\times$ on small generators to $2.9\times$ on 7B models. Future work includes other languages, function-level granularity, and learned routing signals. Agreement and reverse-context likelihood are language-agnostic, and only the dataflow context needs per-language tooling.

\section*{Data Availability}

\noindent A replication package (the \method{} implementation, all prompt builders, the evaluation harness, and the raw predictions and per-task scores for all six models and three benchmarks) is available at \url{https://figshare.com/s/7ef6473e17cec30bcae2}, sufficient to regenerate every table and figure. Upon acceptance, we will release a public, permanently archived version (code, data, and predictions) under an open-source license.

\balance
\bibliographystyle{IEEEtranN}
\bibliography{sections/bibliography}

\end{document}